\documentclass[11pt,english]{article}
\usepackage[T1]{fontenc}
\usepackage{amssymb}
\usepackage{amsfonts}
\usepackage{graphicx}
\usepackage{babel}
\usepackage{a4wide}

\makeatletter
\providecommand{\LyX}{L\kern-.1667em\lower.25em\hbox{Y}\kern-.125emX\@}

\newcommand{\sE}{\mathsf{E}}

\newcommand{\R}{\mathbf{R}}

\newcommand{\btheo}{\begin{theorem}}
\newcommand{\etheo}{\end{theorem}}
\newcommand{\blem}{\begin{lemma}}
\newcommand{\elem}{\end{lemma}}
\newtheorem{theorem}{Theorem}

\newtheorem{corollary}[theorem]{Corollary}

\newtheorem{definition}[theorem]{Definition}

\newtheorem{lemma}[theorem]{Lemma}

\newtheorem{proposition}[theorem]{Proposition}
\newtheorem{remark}[theorem]{Remark}

\makeatother

\begin{document}
\date{\normalsize\null\hspace*{1.64em}Date: \ \quad manuscript --- July 2004,\newline
\null\hspace*{0.5em}ArXiv --- January 2012 \ \ \hspace*{1.08em}\null}

\title{Asymptotic behaviour in the time synchronization model}
\author{V. Malyshev\thanks{ Postal address: Faculty of Mathematics and Mechanics, Moscow State
University, Leninskie Gory 1, \mbox{GSP-1},  119991, Moscow, Russia. \quad
E-mail: \quad malyshev2@yahoo.com, \quad manita@mech.math.msu.su~.\newline 
\hspace*{1\parindent} 
First published in  \   Representation Theory, Dynamical Systems, and Asymptotic Combinatorics,\newline 
 AMS Translations --- Series 2: Advances in the Mathematical Sciences (2006), Vol.~217, pp.~101-115 
}\and  A. Manita}

\maketitle

\vspace*{-2ex}
\begin{abstract}

There are two types \ $i=1,2$ of particles on the line $\R$, with 
$N_{i}$ particles of type~$i\,$. Each particle of type $i$ moves with 
constant velocity $v_{i}$. Moreover, any particle of type $i=1,2$ jumps 
to any particle of type $j=1,2$ with rates $N_{j}^{-1}\alpha _{ij}$. We 
discuss in details the initial desynchronization of this particle 
system, namely, we are interested in behaviour of the process  when the 
total number of particles $N_1+N_2$ tends to infinity, 
$N_1/N_2\rightarrow const$ and the time $t>0$ is fixed.

\medskip 
{\bf Keywords:} Markov process, stochastic particles system,
synchronization model

{\bf 2000 MSC:} 60K35, 60J27, 60F99

\end{abstract}

\section{The Model}

The simplest formulation of the model, we consider here, is in terms of the
particle system. \ On the real line there are $N_{1}$ particles of type $1$
and $N_{2}$ particles of type $2$, $N=N_{1}+N_{2}$. Each particle of type $%
i=1,2$ performs two independent movements. First of all, it moves with
constant speed $v_{i}$ in the positive direction. \ We assume further that $%
v_{i}$ are constant and different, thus we can assume without loss of
generality that $0\leq v_{1}<v_{2}$. The degenerate case $v_{1}=v_{2}$ is
different and will be considered separately.

Secondly, at any time interval $\left[t,t+dt\right]$ each particle of type $%
i $ independently of the others with probability $\alpha _{ij}dt$ decides to
make a jump to some particle of type $j$ and chooses the coordinate of the $%
j $-type particle, where to jump, among the particles of type $j$, with
probability $\frac{1}{N_{j}}$. Here $\alpha _{ij}$ are given nonnegative
parameters for $i,j=1,2$. Further on, unless otherwise stated, we assume
that $\alpha _{11}=\alpha _{22}=0,\quad \alpha _{12},\alpha _{21}>0$.

After such instantaneous jump the particle of type $i$ continues the
movement with the same velocity $v_{i}$. \ This defines continuous time
Markov chain $\left\{ x_{k}^{(i)}(t)\right\} $, $i=1,2$; $k=1,\ldots,N_{i}$, where $%
x_{k}^{(i)}(t)$ is the coordinate of $k$-th particle of type $i$ at time $t$%
. We assume that the initial coordinates $x_{k}^{(i)}(0)$ of the particles
at time $0$ are given. We are interested in the long time evolution of this
system on various scales with $N\rightarrow \infty $, $t=t(N)\rightarrow \infty
$.

In different terms, this can be interpreted as the time synchronization
problem. In general, time synchronization problem can be presented as
follows. There are $N$ systems (processors, units, persons etc.) There is an
absolute (physical) time $t$, but each processor $j$ fulfills a~homogeneous
job in its own proper time $t_{j}=v_{j}t,v_{j}>0$. Proper time is measured
by the amount $v_{j}$ of the job, accomplished by the processor for the unit
of the physical time, if it is disjoint from other processors. However,
there is a communication between each pair of processors, which should lead
to drastic change of their proper times. In our case the coordinates $%
x_{k}^{(i)}(t)$ can be interpreted as the modified proper times of the
particles-processors, the nonmodified proper time being $%
x_{k}^{(i)}(0)+v_{i}t$.

There can be many variants of exact formulation of such problem, see \cite%
{GrMaPo,ManSch,MitMit,BertTsit,MaSi}. We will call the model considered here the basic
model, because there are no restrictions on the jump process. Many other
problems include such restrictions, for example, only jumps to the left are
allowed. Due to absence of restrictions, this problem, as we will see below,
is a "linear problem" in the sense that after scalings it leads to linear
equations. In despite of this it has nontrivial behaviour, one sees
different picture on different time scales.

There are, however, other interesting interpretations of this model, related
to psychology, biology and physics; For example, in social psychology
perception of time and life tempo strongly depends on the social contacts
and intercourse. We will not enter the details here.

\section{Main results}

We show that the process consists of three consecutive stages: initial
desynchronization up to the critical scale, critical slow down of
desynchronization and final stabilization.

\paragraph{Final stabilization}

The first theorem shows that for $N_{i}$ fixed and $t\rightarrow \infty $
there is a synchronization: all particles asymptotically, as $t\rightarrow
\infty $, move with the same constant velocity $v$, that is like $vt$.
However it does not say how fluctuations depend on $N_{i}$.

Put
\[
m(t)=\min _{i,k}x_{k}^{(i)}(t)
\]

\begin{theorem}
\label{t:x/t=3Dv}For any fixed $N_{1},N_{2}$ there exists $%
v=v(N_{1},N_{2})>0 $ such that for any $i=1,2$ and any $k=1,...,N_{i}$ a.s.
\[
\lim _{t\rightarrow \infty }\frac{x_{k}^{(i)}(t)}{t}=v
\]
Moreover, the distribution of the vector $\left\{
x_{k}^{(i)}(t)-m(t),\;i=1,2;\; k=1,\ldots,N_{i}\right\} $ tends to a stationary
distribution.
\end{theorem}

The velocity $v$ will be written down explicitely in terms of this
distribution, it depends of course on $\alpha _{ij}$ and $v_{i}$. Note that
both the velocity and the distribution do not depend on the initial
coordinates.

\paragraph{Initial desynchronization}

Now we consider the case when $N\rightarrow \infty $ but $t$ is fixed. More
exactly, we consider a sequence of pairs $(N_{1},N_{2})$ such that $%
N_{1},N_{2}\rightarrow \infty $ so that $\frac{N_{i}}{N}\rightarrow c_{i}$,
where $c_{1}+c_{2}=1,c_{i}>0$. It is convenient here to consider positive
measures or generalized functions%
\[
m^{(N_{i})}(t,x)=\frac{1}{N_{i}}\sum_{k}\delta (x-x_{k}^{(i)}(t)),\quad x\in \R _{+}
\]%
defined by the coordinates of $N_{i}$ particles of type $i$ at time $t$. We
assume that at time $t=0$ for any bounded $C^{1}$-functions $\phi _{i}(x)$
on $\R $ the sequence $<m_{i}^{(N_{i})}(0,.),\phi _{i}>$ converges to some
number.

\begin{theorem}
\label{t:N-infinity-t-fixed}Then for any $t$ there are weak deterministic
limits%
\[
\lim _{N\rightarrow \infty }\frac{1}{N}m_{i}^{(N_{i})}(t,x)=m_{i}(t,x)
\]
where $m_{i}(t,x)$ satisfy the following equations%
\begin{equation}
\frac{\partial m_{1}}{\partial t}+v_{1}\frac{\partial m_{1}}{\partial x}%
=\alpha _{12}(m_{2}-m_{1})  \label{cont_1}
\end{equation}
\begin{equation}
\frac{\partial m_{2}}{\partial t}+v_{2}\frac{\partial m_{2}}{\partial x}%
=\alpha _{21}(m_{1}-m_{2})  \label{cont_2}
\end{equation}
\end{theorem}

Now we want to study the asymptotic behaviour of $m_{i}(x,t)$ for $%
t\rightarrow \infty $. Denote%
\[
a_{i}(t)=\int xm_{i}(x,t)dx,\quad \quad d_{i}(t)=\int
(x-a_{i}(t))^{2}m_{i}(t,x)\, dx
\]

\begin{theorem}
\label{t-a-d-cont}There exist constants $v,d>0$ such that as $t\rightarrow
\infty $%
\[
a_{i}(t)=vt+a_{i0}+o(1)
\]%
\[
d_{i}(t)=dt+d_{i0}+o(1)
\]%
for some constants $a_{i0},d_{i0}$. Moreover,%
\[
\Delta _{i}(x,t)=\frac{m_{i}(x,t)-a_{i}(t)}{\sqrt{d_{i}(t)}}
\]%
tends to $\frac{1}{\sqrt{2\pi }}\exp (-\frac{x^{2}}{2})$ pointwise as $%
t\rightarrow \infty $.
\end{theorem}

\paragraph{Critical point and uniform estimates}

Here we assume that $N_{1}=[c_{1}N]$, $N_{2}=[c_{2}N]$ for some $%
c_{i}>0,c_{1}+c_{2}=1$. Introduce the empirical means (mass centres) for
types 1 and 2%
\[
\overline{x^{(i)}}(t)=\frac{1}{N_{i}}\sum_{k=1}^{N_{i}}x_{k}^{(i)}(t),
\]%
the empirical variances
\[
S_{i}^{2}(t)=\frac{1}{N_{i}}\sum_{k=1}^{N_{i}}\left( x_{k}^{(i)}(t)-%
\overline{x^{(i)}}(t)\right) ^{2}
\]%
and their means%
\[
\mbox{\boldmath$\mu$}_{i}(t)=\mathsf{E}\overline{x^{(i)}}(t),\quad %
\mbox{\boldmath $l$}_{12}(t)=\mbox{\boldmath$\mu$}_{1}(t)-%
\mbox{\boldmath$\mu$}_{2}(t),\quad R_{i}(t)=\mathsf{E}S_{i}^{2}(t)
\]%
The following asymptotic results hold for any sequence of pairs $(N,t)$ with
$N\rightarrow \infty $ and $t=t(N)\rightarrow \infty $.

\begin{theorem}
\label{t:l-mu}We have the following asymptotical results
as $t\rightarrow \infty $:
\[
\mbox{\boldmath $l$}_{12}(t)\rightarrow \frac{v_{1}-v_{2}}{\alpha
_{12}+\alpha _{21}},\qquad \frac{\mbox{\boldmath$\mu$}_{i}(t)}{t}\rightarrow
\frac{\alpha _{12}v_{2}+\alpha _{21}v_{1}}{\alpha _{12}+\alpha _{21}}
\]
\end{theorem}

Assume now that $N_{i}=c_{i}N$, where $c_{i}>0,c_{1}+c_{2}=1$.

\begin{theorem}
\label{t-RR} There are the following three regions of asymptotic behaviour,
uniform in $t(N)$ for sufficiently large $N$:

\begin{itemize}
\item if ${\displaystyle\frac{t(N)}{N}\rightarrow 0}$, then $R_{i}(t(N))\sim
h\varkappa _{2}t(N)$,

\item if $t=t(N)=sN$ for some $s>0$, then $R_{i}(t(N))\sim
h\,(1-e^{-\varkappa _{2}s})N$,

\item if ${\displaystyle\frac{t(N)}{N}\rightarrow \infty }$, then $%
R_{i}(t(N))\sim hN$,
\end{itemize}

where the constant $\varkappa _{2}>0$ can be explicitely calculated and
\[
h=\frac{2\alpha _{12}\alpha _{21}\left( v_{1}-v_{2}\right) ^{2}}{\varkappa
_{2}(\alpha _{12}+\alpha _{21})^{3}}\,.
\]
\end{theorem}
The proofs of Theorems \ref{t:l-mu} and \ref{t-RR} are given in~\cite{MalMan1,MalMan2}.

\section{Limit $t\rightarrow \infty $}

In this section we will prove Theorem~\ref{t:x/t=3Dv}.

\paragraph{Two particles.}

It is useful to consider first the case when $N_{1}=N_{2}=1$. Thus consider
the process $(x^{(1)}(t),x^{(2)}(t))$. We will prove that there exist
deterministic limits%
\[
\lim_{t\rightarrow \infty }\frac{x^{(i)}(t)}{t}=v
\]%
for $i=1,2$ and some $v>0$, moreover the distribution of the random variable
$\rho (t)=x^{(2)}(t)-x^{(1)}(t)$ tends to some distribution on $\R _{+}$.

We can assume that $v_{1}=0,v_{2}>0$. The Markov chain $\rho
(t)=x^{(2)}(t)-x^{(1)}(t)$ on $\R _{+}$ satisfies the Doeblin condition, that
is from any $x\in \R _{+}$ there is a jump rate to $0$, bounded away from
zero, here it equals $\alpha _{12}+\alpha _{21}$. It follows that $\rho (t)$
is ergodic. Then as $t\rightarrow \infty $ there exists the limiting
(invariant) distribution $F(x)$ for $\rho (t)$. Let%
\[
t_{1}<t_{2}< \cdots
\]
time moments when $x^{(1)}(t)=x^{(2)}(t)$. It is clear that $t_{k}-t_{k-1}$
are independent random variables, exponentially distributed with parameter $%
\alpha _{12}+\alpha _{21}$. It follows that $F(x)$ is exponential with the
density%
\[
p(x)=\lambda \exp (-\lambda x),\quad \lambda =\frac{\alpha _{12}+\alpha _{21}%
}{v_{2}-v_{1}}
\]

Thus, if the limits $\lim\limits _{t\rightarrow \infty }\frac{x_{i}(t)}{t}$\ exist,
then they are equal. Let us prove that they exist and
\begin{equation}
v=v_{1}{}+\alpha _{12}\int xp(x)dx  \label{v-1}
\end{equation}
In fact, the particle $1$ moves with constant speed $v_{1}$ and performs on
the time interval $\left[0,T\right]$ independent exponentially distributed
jumps in the positive direction. As $T\rightarrow \infty $, the number of
these jumps asymptotically equals $\alpha _{12}T$, and the mean jump
asymptotically is $\int xp(x)dx$.

Similarly one can get%
\begin{equation}
v=v_{2}{}-\alpha _{21}\int xp(x)dx  \label{v-2}
\end{equation}
From this and (\ref{v-1}) we have%
\[
v=\frac{\alpha _{21}v_{1}+\alpha _{12}v_{2}}{\alpha _{21}+\alpha _{12}}
\]

\paragraph{General case.}

Let us prove first the second statement of the theorem. We can put $v_{1}=0$
and change the coordinate system putting $m(t)=0$. Consider a configuration
of particles at time $t$. Denote the particle, which has coordinate $m(t)=0$
at time $t$, as particle $0$. Let $p(t+2)$ be the probability that at time $%
t+2$ each particle will be inside the interval $\left[0,2v_{2}\right]$. This
probability can be (very roughly) estimated from \ below as%
\[
p(t+2)\geq \min (p_{01}p_{2}p_{1},p_{02}p_{3}p_{4})
\]
To prove this consider first the case when the particle $0$ has type $1$.
Under this condition $p(t+2)$ can be estimated from below as $%
p_{01}p_{2}p_{1}$, where $p_{01}$ is the probability that particle $0$ does
not do any jumps in the time interval $(t,t+2)$, $p_{2}$ is the probability
that each particle of type $2$ jumps at least once to the particle $0$ in
the time interval $(t,t+1)$ and does not do any more jumps in the time
interval $(t,t+2)$, $p_{1}$ is the probability that each particle of type $1$
jumps to some particle of type $2$ in the time interval $(t+1,t+2)$.
Similarly, under the condition that the particle $0$ has type $2$, $p(t+2)$
can be estimated from below as $p_{02}p_{3}p_{4}$, where $p_{02}$ is the
probability that the particle $0$ does not do any jumps in the time interval
$(t,t+2)$, $p_{3}$ is the probability that each particle of type $1$ jumps
at least once to the particle $0$ in the time interval $(t,t+1)$ and does
not do any more jumps in the time interval $(t,t+2)$, $p_{4}$ is the
probability that each particle of type $2$ jumps to some particle of type $1$
in the time interval $(t+1,t+2)$.

This means that the Markov chain $\mathcal{L}=\left\{
x_{k}^{(i)}(t)-m(t),i=1,2;k=1,...,N_{i}\right\} $ satisfies the Doeblin
condition. Then it is ergodic and has some stationary distribution. We will
write now formula for $v$, assuming however that $\alpha _{ii}=0$. For this
we need some marginals of this stationary distribution.

Let $A_{i}(t)$ be the event that at time $t$ at the point $m(t)$ there is a
particle of type $i$, and $q_{i}=\lim \limits_{t\rightarrow \infty }P(A_{i}(t))$
be the stationary (limiting) probability of $A_{i}$. Let $p_{i}(y)$ be the
stationary conditional (under the condition $A_{i}$) probability density of
the distance from $m$ to the nearest particle. In the time interval $%
[T,T+dt] $ the particle in $m(t)$ moves with the speed $v_{i}$, and moreover
can make one jump. This gives, for example under the condition $A_{1}$,
constant movement $v_{1}dt$ of $m$, and the jump of $m$ to the nearest point
with rate $\alpha _{12}dt$. Thus as $T\rightarrow \infty $ we have
\begin{eqnarray*}
\mathsf{E} (m(T+dt)\, |\, m(T)\, )\, -\, m(T)\, &=&\,
q_{1}\,\Bigl(v_{1}+\alpha_{12}\int yp_{1}(y)dy\Bigr)\,dt\,+{ } \\
&&
q_{2}\,\Bigl(v_{2}+\alpha _{21}\int yp_{2}(y)dy\Bigr)\,dt\,+\, o(1)
\end{eqnarray*}
and then%
\[
v=q_{1}\,\Bigl(v_{1}+\alpha _{12}\int yp_{1}(y)dy\Bigr)+
q_{2}\,\Bigl(v_{2}+\alpha _{21}\int
yp_{2}(y)dy\Bigr)
\]

\paragraph{About Doeblin chains.}

In the standard theory of Doeblin chains, see \cite{Doe}, it is assumed that
transition probabilities are absolutely continuous with respect to some
positive measure $\mu $ on the state space.

If at time $0$ all $x_{k}^{(i)}$ are different, then for any $t$ it is true
that all $x_{k}^{(i)}$ are different a.s. Thus transition probabilities (for
example, for the embedded chain at times $0,1,2,\ldots$) are absolutely
continuous with respect to Lebesgue measure on $(\R _{+}^{N_{1}-1}\times
\R _{+}^{N_{2}})\cup (\R _{+}^{N_{1}}\times \R _{+}^{N_{2}-1})$. If at
time~$0$
some coordinates coincide, then a.s. in finite time $\tau $ they become all
different.

\section{Limit $N\rightarrow \infty $}

It is very intuitive to introduce the following continuous model. \ Let $%
m_{i}(0,x)$, $x\in \R $, $i=1,2,$ be positive smooth functions, $M_{i}=\int
m_{i}(0,x)dx=1$. We call them continuous mass distributions of type $i$ at
time $t=0$. The dynamics of the masses is deterministic --- during time $dt$
from each element $dm_{1}$ of the mass the part $\alpha _{12}\,dt\,dm_{1}$ goes
out and distributes correspondingly to the mass $m_{2}(x)$, namely it
becomes the mass distribution with density $m_{2}(x)\,\alpha _{12}\,dt\,dm_{1}$,
and vice-versa, interchanging $1$ and $2$. Moreover each mass element moves
with velocities $v_{1}$ and $v_{2}$ correspondingly. From this we easily get
linear equations (\ref{cont_1})--(\ref{cont_2}) for mass distribution $%
m_{i}(t,x)$ at time $t$ with the initial conditions%
\[
m_{i}(0,x)=f_{i}(x)
\]

Now we will prove convergence of $N$ particle model to the continuous model.

\subsection{Convergence: the martingale problem}

Here we prove Theorem~\ref{t:N-infinity-t-fixed}.

We consider continuous time Markov process%
\begin{equation}
\xi _{N_{1},N_{2}}(t)=\left(x_{1}^{(1)}(t),\ldots
,x_{N_{1}}^{(1)}(t);x_{1}^{(2)}(t),\ldots ,x_{N_{2}}^{(2)}(t)\right)
\label{eq:xiNN-t}
\end{equation}
with the state space~$\R ^{N_{1}+N_{2}}$. Its generator
\begin{eqnarray*}
\left(L_{N_{1},N_{2}}f\right)\left(x_{\, }^{(1)};x_{_{\, }}^{(2)}\right) & =
& \left[v_{1}\sum _{i=1}^{N_{1}}\frac{\partial }{\partial x_{i}^{(1)}}%
+v_{2}\sum _{j=1}^{N_{2}}\frac{\partial }{\partial x_{j}^{(2)}}\right]%
f\left(x_{\, }^{(1)};x_{_{\, }}^{(2)}\right)+ \\
& & +\sum _{i=1}^{N_{1}}\sum _{j=1}^{N_{2}}\frac{\alpha _{12}}{N_{2}}\left[%
f\left(\left(x_{\, }^{(1)};x_{_{\, }}^{(2)}\right)_{i\rightarrow
j}\right)-f\left(x_{\, }^{(1)};x_{_{\, }}^{(2)}\right)\right]+ \\
& & +\sum _{j=1}^{N_{2}}\sum _{i=1}^{N_{1}}\frac{\alpha _{21}}{N_{1}}\left[%
f\left(\left(x_{\, }^{(1)};x_{_{\, }}^{(2)}\right)_{i\leftarrow
j}\right)-f\left(x_{\, }^{(1)};x_{_{\, }}^{(2)}\right)\right],
\end{eqnarray*}
where the following notation is used
\begin{eqnarray*}
\left(x_{\, }^{(1)};x_{_{\, }}^{(2)}\right) & = & \left(x_{1}^{(1)},\ldots
,x_{N_{1}}^{(1)};x_{1}^{(2)},\ldots ,x_{N_{2}}^{(2)}\right), \\
\left(x_{\, }^{(1)};x_{_{\, }}^{(2)}\right)_{i\rightarrow j} & = &
\left(x_{1}^{(1)},\ldots ,x_{i-1}^{(1)},x_{j}^{(2)},x_{i+1}^{(1)},\ldots
,x_{N_{1}}^{(1)};x_{1}^{(2)},\ldots ,x_{N_{2}}^{(2)}\right), \\
\left(x_{\, }^{(1)};x_{_{\, }}^{(2)}\right)_{i\leftarrow j} & = &
\left(x_{1}^{(1)},\ldots ,x_{N_{1}}^{(1)};x_{1}^{(2)},\ldots
,x_{j-1}^{(2)},x_{i}^{(1)},x_{j+1}^{(2)},\ldots ,x_{N_{2}}^{(2)}\right),
\end{eqnarray*}
is defined on bounded $C^{1}$-functions.

We will consider the limiting behaviour of this process when $t=const$, $%
N_{1},N_{2}\rightarrow \infty $. It is not convenient to deal with the
sequence $\xi _{N_{1},N_{2}}(t)$ of processes because the dimension of the
state space changes with $N_{1},N_{2}$.

Denote
\[
M_{N_{1},N_{2}}(t)=\left( \frac{1}{N_{1}}\sum_{i=1}^{N_{1}}\delta (\cdot
-x_{i}^{(1)}(t)),\frac{1}{N_{2}}\sum_{j=1}^{N_{2}}\delta (\cdot
-x_{j}^{(2)}(t))\right) .
\]%
where $\delta (x),\,x\in \R ,$ is the $\delta $-function. One can see that the
generalized functions
\[
\frac{1}{N_{1}}\sum_{i=1}^{N_{1}}\delta (\cdot -x_{i}^{(1)}(t)),\qquad \frac{%
1}{N_{2}}\sum_{j=1}^{N_{2}}\delta (\cdot -x_{j}^{(2)}(t))
\]%
represent empirical "densities" or masses of (type $1$ and
$2$ correspondingly) particles at time~$t$. Thus, if $\phi (x)=(\phi
_{1}(x),\phi _{2}(x))$, where $\phi _{i}\in S(\R )$, then for fixed
particle positions $%
x_{1}^{(1)}(t),\ldots ,x_{N_{1}}^{(1)}(t)$ and $x_{1}^{(2)}(t),\ldots
,x_{N_{2}}^{(2)}(t)$ the vector function $M_{N_{1},N_{2}}(t)$ is a linear
functional on the vector test functions~$\phi $, that is
\[
\left\langle M_{N_{1},N_{2}}(t),\phi \right\rangle =\frac{1}{N_{1}}%
\sum_{i=1}^{N_{1}}\phi _{1}(x_{i}^{(1)}(t))+\frac{1}{N_{2}}%
\sum_{j=1}^{N_{2}}\phi _{2}(x_{j}^{(2)}(t)).
\]

Fix some $T>0$. Then $\left( M_{N_{1},N_{2}}(t),\,0\leq t\leq T\right) $ can
be considered as a Markov process taking its values in the space of tempered
distributions $S^{\prime }(\R )\times S^{\prime }(\R )$. In the sequel we
consider $S^{\prime }(\R )\times S^{\prime }(\R )$ as a topological space
equipped with the strong topology (see Subsection~\ref%
{sub:Strong-topology-on}). Without loss of generality one can assume that
the trajectories of the process $M_{N_{1},N_{2}}(t)$ are right continuous
functions with left limits. So it is natural to consider the Skorohod space $%
\Pi ^{T}=D([0,T],S^{\prime }(\R )\times S^{\prime }(\R ))$ of functions on $%
[0,T] $ with values in $S^{\prime }(\R )\times S^{\prime }(\R )$ as a coordinate
space of the process~$M_{N_{1},N_{2}}(t)$. Subsection~\ref%
{sub:Strong-topology-on} explains how to introduce topology on this space.
Let $\mathcal{B}(\Pi ^{T})$ be the corresponding Borel $\sigma $-algebra.
Denote $P_{N_{1},N_{2}}^{T}$ the probability measure on $\left( \Pi ^{T},%
\mathcal{B}(\Pi ^{T})\right) $, induced by the process $\left(
M_{N_{1},N_{2}}(t),\,0\leq t\leq T\right) $.

Our assumption for the theorem is that for any test function $\phi (x)$ the
sequence $\left\langle M_{N_{1},N_{2}}(0),\phi \right\rangle $ weakly
converges as $N_{1},N_{2}\rightarrow \infty $.

We want to prove that as $N_{1},N_{2}\rightarrow \infty $ the sequence of
probability distributions $P_{N_{1},N_{2}}^{T}$ has a weak limit, and this
limit is a one-point measure, that is the only trajectory $%
(m_{1}(t),m_{2}(t))$, $0\leq t\leq T$, which is the classical solution of
the system (\ref{cont_1})-(\ref{cont_2}). We split a proof of this result
into the next two propositions.

\begin{proposition}
\label{p:family-dense}The family of probability distributions $\left\{
P_{N_{1},N_{2}}^{T}\right\} _{N_{1},N_{2} }$ on $\left(\Pi ^{T},\mathcal{B}%
(\Pi ^{T})\right)$ is tight.
\end{proposition}

\begin{proposition}
\label{p-Limit-points}Limit points of the family of distributions $%
P_{N_{1},N_{2}}^{T}$ are concentrated on the weak solutions of the system~(%
\ref{cont_1})-(\ref{cont_2}).
\end{proposition}

\subsubsection{Tightness}

Before proving Proposition~\ref{p:family-dense} we start with some
preliminary lemmas. We want to prove that the family of distributions $%
P_{N_{1},N_{2}}^{T}$ of the random process $\left(
M_{N_{1},N_{2}}(t),\,0\leq t\leq T\right) $, with values in the space of
generalized functions, is tight. By the theorem 4.1 of~\cite{mitoma} (see
also Subsection~\ref{sub:Strong-topology-on}), it is sufficient to prove
that for any test function $\psi =\left( \psi _{1}(x),\psi _{2}(x)\right) $
the family of random processes $\left( \left\langle M_{N_{1},N_{2}}(t),\psi
\right\rangle ,\,0\leq t\leq T\right) $, with values in $\R ^{1}$, is tight.
This will be done in the Proposition~\ref{e-plot-Fpsi} below.

Fix some test function $\psi =\left(\psi _{1}(x),\psi _{2}(x)\right)$ and
consider the random process%
\begin{eqnarray*}
F_{\psi ,N_{1},N_{2}}\left(x_{\, }^{(1)}(t);x_{_{\, }}^{(2)}(t)\right) & = &
\left\langle M_{N_{1},N_{2}}(t),\psi \right\rangle = \\
& = & \frac{1}{N_{1}}\sum _{i=1}^{N_{1}}\psi _{1}(x_{i}^{(1)}(t))+\frac{1}{%
N_{2}}\sum _{j=1}^{N_{2}}\psi _{2}(x_{j}^{(2)}(t))
\end{eqnarray*}
This is a function of the Markov process $\xi _{N_{1},N_{2}}(t)$, thus (see~%
\cite[Lemma 5.1, p.~330]{KipnLand}, for example) the following two processes
are martingales:%
\begin{eqnarray}
W_{\psi ,N_{1},N_{2}}(t) & = & F_{\psi ,N_{1},N_{2}}\left(x_{\,
}^{(1)}(t);x_{_{\, }}^{(2)}(t)\right)-F_{\psi ,N_{1},N_{2}}\left(x_{\,
}^{(1)}(0);x_{_{\, }}^{(2)}(0)\right)-  \nonumber \\
& & -\int _{0}^{t}L_{N_{1},N_{2}}F_{\psi ,N_{1},N_{2}}\left(x_{\,
}^{(1)}(s);x_{_{\, }}^{(2)}(s)\right)ds  \label{eq:W-repr} \\
V_{\psi ,N_{1},N_{2}}(t) & = & \left(W_{\psi
,N_{1},N_{2}}(t)\right)^{2}-\int _{0}^{t}L_{N_{1},N_{2}}F_{\psi
,N_{1},N_{2}}^{2}\left(x_{\, }^{(1)}(s);x_{_{\, }}^{(2)}(s)\right)ds+
\nonumber \\
& & +2\int _{0}^{t}F_{\psi ,N_{1},N_{2}}\left(x_{\, }^{(1)}(s);x_{_{\,
}}^{(2)}(s)\right)L_{N_{1},N_{2}}F_{\psi ,N_{1},N_{2}}\left(x_{\,
}^{(1)}(s);x_{_{\, }}^{(2)}(s)\right)ds.  \nonumber
\end{eqnarray}
For shortness we will write $F(x_{\, }^{(1)};x_{_{\, }}^{(2)})$ instead of $%
F_{\psi ,N_{1},N_{2}}(x_{\, }^{(1)};x_{_{\, }}^{(2)})$.

\begin{lemma}
The following estimates hold:

\begin{description}
\item[i)] $\left|L_{N_{1},N_{2}}F\left(x_{\, }^{(1)};x_{_{\,
}}^{(2)}\right)\right|\leq C_{1}(\psi ,v_{1},v_{2},\alpha _{12},\alpha
_{21}) $ uniformly in $N_{1},N_{2}$ and $\left(x_{\, }^{(1)};x_{_{\,
}}^{(2)}\right) $;

\item[ii)] uniformly in $x_{\, }^{(1)},\, x_{_{\, }}^{(2)}$%
\begin{equation}
\left|L_{N_{1},N_{2}}F^{2}(x_{\, }^{(1)};x_{_{\, }}^{(2)})-F(x_{\,
}^{(1)};x_{_{\, }}^{(2)})L_{N_{1},N_{2}}F(x_{\, }^{(1)};x_{_{\,
}}^{(2)})\right|\leq \frac{C_{12}(\alpha _{12},\psi _{1})}{N_{1}}+\frac{%
C_{21}(\alpha _{21},\psi _{2})}{N_{2}}.  \label{eq:LF2}
\end{equation}
\end{description}
\end{lemma}

\noindent \textbf{Proof of the lemma.} Note that%
\begin{eqnarray*}
F\left( \left( x_{\,}^{(1)};x_{_{\,}}^{(2)}\right) _{i\rightarrow j}\right)
-F\left( x_{\,}^{(1)};x_{_{\,}}^{(2)}\right)  &=&\frac{1}{N_{1}}\left( \psi
_{1}\left( x_{j}^{(2)}\right) -\psi _{1}\left( x_{i}^{(1)}\right) \right) ,
\\
F\left( \left( x_{\,}^{(1)};x_{_{\,}}^{(2)}\right) _{i\leftarrow j}\right)
-F\left( x_{\,}^{(1)};x_{_{\,}}^{(2)}\right)  &=&\frac{1}{N_{2}}\left( \psi
_{2}\left( x_{i}^{(1)}\right) -\psi _{2}\left( x_{j}^{(2)}\right) \right) .
\end{eqnarray*}%
Thus%
\begin{eqnarray}
L_{N_{1},N_{2}}F\left( x_{\,}^{(1)};x_{_{\,}}^{(2)}\right)  &=&\frac{v_{1}}{%
N_{1}}\sum_{i=1}^{N_{1}}\psi _{1}^{\prime }\left( x_{i}^{(1)}\right) +\frac{%
v_{2}}{N_{2}}\sum_{j=1}^{N_{2}}\psi _{2}^{\prime }\left( x_{j}^{(2)}\right) +
\nonumber \\
&&+\sum_{i=1}^{N_{1}}\sum_{j=1}^{N_{2}}\frac{\alpha _{12}}{N_{2}}\cdot \frac{%
1}{N_{1}}\left( \psi _{1}\left( x_{j}^{(2)}\right) -\psi _{1}\left(
x_{i}^{(1)}\right) \right) +  \nonumber \\
&&+\sum_{j=1}^{N_{2}}\sum_{i=1}^{N_{1}}\frac{\alpha _{21}}{N_{1}}\cdot \frac{%
1}{N_{2}}\left( \psi _{2}\left( x_{i}^{(1)}\right) -\psi _{2}\left(
x_{j}^{(2)}\right) \right) .  \label{eq:LF-action}
\end{eqnarray}%
Then%
\begin{eqnarray*}
\left\vert L_{N_{1},N_{2}}F\left( x_{\,}^{(1)};x_{_{\,}}^{(2)}\right)
\right\vert  &\leq &\left\vert v_{1}\right\vert \left\Vert \psi _{1}^{\prime
}\right\Vert _{C}+\left\vert v_{2}\right\vert \left\Vert \psi _{2}^{\prime
}\right\Vert _{C}+ \\
&&+2\alpha _{12}\left\Vert \psi _{1}\right\Vert _{C}+2\alpha _{21}\left\Vert
\psi _{2}\right\Vert _{C}\,
\end{eqnarray*}%
and the assertion \textbf{i)} of the lemma is proved. To prove assertion
\textbf{ii)} it is convenient to represent $%
L_{N_{1},N_{2}}=L_{N_{1},N_{2}}^{0}+L_{N_{1},N_{2}}^{1}$ as the sum of \char`%
\"{}differential\char`\"{} $L_{N_{1},N_{2}}^{0}$and \char`\"{}jump\char`\"{}
$L_{N_{1},N_{2}}^{1}$ parts.

It is easy to see that
\[
L_{N_{1},N_{2}}^{0}F^{2}(x_{\, }^{(1)};x_{_{\, }}^{(2)})-2F(x_{\,
}^{(1)};x_{_{\, }}^{(2)})L_{N_{1},N_{2}}^{0}F(x_{\, }^{(1)};x_{_{\,
}}^{(2)})=0.
\]
Let us prove that uniformly in $x_{\, }^{(1)},\, x_{_{\, }}^{(2)}$%
\begin{equation}
\left|L_{N_{1},N_{2}}^{1}F^{2}(x_{\, }^{(1)};x_{_{\, }}^{(2)})-F(x_{\,
}^{(1)};x_{_{\, }}^{(2)})L_{N_{1},N_{2}}^{1}F(x_{\, }^{(1)};x_{_{\,
}}^{(2)})\right|\leq \frac{4\alpha _{12}\left\Vert \psi _{1}\right\Vert
_{C}^{2}}{N_{1}}+\frac{4\alpha _{21}\left\Vert \psi _{2}\right\Vert _{C}^{2}%
}{N_{2}}.  \label{eq:L1F2}
\end{equation}
In fact
\begin{eqnarray*}
F^{2}\left(\left(x_{\, }^{(1)};x_{_{\, }}^{(2)}\right)_{i\rightarrow
j}\right)-F^{2}\left(x_{\, }^{(1)};x_{_{\, }}^{(2)}\right) & = &
\left(F\left(\left(x_{\, }^{(1)};x_{_{\, }}^{(2)}\right)_{i\rightarrow
j}\right)-F\left(x_{\, }^{(1)};x_{_{\, }}^{(2)}\right)\right)\times \\
& & \times \left(2F\left(x_{\, }^{(1)};x_{_{\, }}^{(2)}\right)+\frac{1}{N_{1}%
}\left(\psi _{1}\left(x_{j}^{(2)}\right)-\psi
_{1}\left(x_{i}^{(1)}\right)\right)\right) \\
& = & 2F\left(x_{\, }^{(1)};x_{_{\, }}^{(2)}\right)\left[F\left(\left(x_{\,
}^{(1)};x_{_{\, }}^{(2)}\right)_{i\rightarrow j}\right)-F\left(x_{\,
}^{(1)};x_{_{\, }}^{(2)}\right)\right]+ \\
& & +\left[\frac{1}{N_{1}}\left(\psi _{1}\left(x_{j}^{(2)}\right)-\psi
_{1}\left(x_{i}^{(1)}\right)\right)\right]^{2}.
\end{eqnarray*}
and similarly for expressions with $\left(x_{\, }^{(1)};x_{_{\,
}}^{(2)}\right)_{i\leftarrow j}$. Thus
\begin{eqnarray*}
L_{N_{1},N_{2}}^{1}F^{2}(x_{\, }^{(1)};x_{_{\, }}^{(2)}) & = & 2F\left(x_{\,
}^{(1)};x_{_{\, }}^{(2)}\right)\sum _{i=1}^{N_{1}}\sum _{j=1}^{N_{2}}\frac{%
\alpha _{12}}{N_{2}}\cdot \left[F\left(\left(x_{\, }^{(1)};x_{_{\,
}}^{(2)}\right)_{i\rightarrow j}\right)-F\left(x_{\, }^{(1)};x_{_{\,
}}^{(2)}\right)\right] \\
& & +\sum _{i=1}^{N_{1}}\sum _{j=1}^{N_{2}}\frac{\alpha _{12}}{N_{2}}\cdot %
\left[\frac{1}{N_{1}}\left(\psi _{1}\left(x_{j}^{(2)}\right)-\psi
_{1}\left(x_{i}^{(1)}\right)\right)\right]^{2} \\
& & +2F\left(x_{\, }^{(1)};x_{_{\, }}^{(2)}\right)\sum _{j=1}^{N_{2}}\sum
_{i=1}^{N_{1}}\frac{\alpha _{21}}{N_{1}}\cdot \left[F\left(\left(x_{\,
}^{(1)};x_{_{\, }}^{(2)}\right)_{i\leftarrow j}\right)-F\left(x_{\,
}^{(1)};x_{_{\, }}^{(2)}\right)\right] \\
& & +\sum _{j=1}^{N_{2}}\sum _{i=1}^{N_{1}}\frac{\alpha _{21}}{N_{1}}\cdot %
\left[\frac{1}{N_{2}}\left(\psi _{2}\left(x_{i}^{(1)}\right)-\psi
_{2}\left(x_{j}^{(2)}\right)\right)\right]^{2} \\
& & \\
& = & 2FL_{N_{1},N_{2}}^{1}F+\frac{\alpha _{12}}{N_{1}}\sum
_{i=1}^{N_{1}}\sum _{j=1}^{N_{2}}\frac{1}{N_{2}N_{1}}\left(\psi
_{1}\left(x_{j}^{(2)}\right)-\psi _{1}\left(x_{i}^{(1)}\right)\right)^{2}- \\
& & +\frac{\alpha _{21}}{N_{2}}\sum _{j=1}^{N_{2}}\sum _{i=1}^{N_{1}}\frac{1%
}{N_{1}N_{2}}\left(\psi _{2}\left(x_{i}^{(1)}\right)-\psi
_{2}\left(x_{j}^{(2)}\right)\right)^{2}.
\end{eqnarray*}
the estimate (\ref{eq:L1F2}) follows from this. Lemma is proved.

\begin{corollary}
\label{c-w2}%
\[
\sup _{t\leq T}\mathsf{E}\left(W_{\psi ,N_{1},N_{2}
}(t)\right)^{2}\rightarrow 0,\qquad N_{1},N_{2} \rightarrow \infty .
\]
\end{corollary}

\noindent \textbf{Proof.} As $V_{\psi ,N_{1},N_{2}}$ is a martingale with
mean zero, it is sufficient to prove that the expectation of%
\[
\int _{0}^{t}\left[L_{N_{1},N_{2}}F^{2}\left(x_{\, }^{(1)}(s);x_{_{\,
}}^{(2)}(s)\right)-2F\left(x_{\, }^{(1)}(s);x_{_{\,
}}^{(2)}(s)\right)L_{N_{1},N_{2}}F\left(x_{\, }^{(1)}(s);x_{_{\,
}}^{(2)}(s)\right)\right]ds
\]
tends to zero. This follows from the estimate~(\ref{eq:LF2}) of the lemma.

\begin{proposition}
\label{e-plot-Fpsi}

The sequence of distributions of real valued random processes
\[
F_{\psi ,N_{1},N_{2}}\left( x_{\,}^{(1)}(t);x_{_{\,}}^{(2)}(t)\right),\;t\in[%
0,T],
\]
is tight.
\end{proposition}

\noindent \textbf{Proof of Proposition~\ref{e-plot-Fpsi}.} Remind that the
following representation holds%
\begin{eqnarray*}
F_{\psi ,N_{1},N_{2}}\left(x_{\, }^{(1)}(t);x_{_{\, }}^{(2)}(t)\right) & = &
F_{\psi ,N_{1},N_{2}}\left(x_{\, }^{(1)}(0);x_{_{\,
}}^{(2)}(0)\right)+W_{\psi ,N_{1},N_{2}}(t)+ \\
& & +\int _{0}^{t}L_{N_{1},N_{2}}F_{\psi ,N_{1},N_{2}}\left(x_{\,
}^{(1)}(s);x_{_{\, }}^{(2)}(s)\right)ds
\end{eqnarray*}
Note that our initial assumption is that the sequence $F_{\psi
,N_{1},N_{2}}\left(x_{\, }^{(1)}(0);x_{_{\, }}^{(2)}(0)\right)$ weakly
converges as $N_{1},N_{2}\rightarrow \infty $.

Prove now that the sequence
\[
\left\{ \eta ^{N_{1},N_{2}}(t)=\int_{0}^{t}L_{N_{1},N_{2}}F\left(
x_{\,}^{(1)}(s);x_{_{\,}}^{(2)}(s)\right) ds,\,t\in \lbrack 0,T]\right\}
_{N_{1},N_{2}}\,.
\]%
is tight. We use subsection 6.1 of the Appendix. By assertion \textbf{i)} of
the lemma
\[
\left\vert \int_{0}^{t}L_{N_{1},N_{2}}F\left(
x_{\,}^{(1)}(s);x_{_{\,}}^{(2)}(s)\right) ds\right\vert \leq C_{1}(\psi
,v_{1},v_{2},\alpha _{12},\alpha _{21})\cdot T\,,
\]%
thus, the condition 1) of the Appendix holds. The condition 2) also holds,
as one can prove that
\[
w^{\prime }(\eta ^{N_{1},N_{2}},\gamma )\leq 2\gamma \cdot C_{1}(\psi
,v_{1},v_{2},\alpha _{12},\alpha _{21})\,.
\]

Prove that the sequence $\left\{ W_{\psi ,N_{1},N_{2}}(t),\,t\in \lbrack
0,T]\right\} _{N_{1},N_{2}}$ is tight. Using Kolmogorov's inequality for
submartingales with right continuous trajectories (see \cite{Doe}), we have
the following estimate, uniform in $N_{1},N_{2}$,
\[
P\left( \sup_{t\leq T}\left\vert W_{\psi ,N_{1},N_{2}}(t)\right\vert
>C\right) \leq \frac{\sup_{t\leq T}\mathsf{E}\left( W_{\psi
,N_{1},N_{2}}(t)\right) ^{2}}{C^{2}}\,
\]%
Then from the corollary~\ref{c-w2} the condition 1) of Appendix holds. Thus%
\begin{eqnarray*}
P\left( \left\vert W_{\psi ,N_{1},N_{2}}(\tau +\theta )-W_{\psi
,N_{1},N_{2}}(\tau )\right\vert >\varepsilon \right) &\leq &\frac{\mathsf{E}%
\left( W_{\psi ,N_{1},N_{2}}(\tau +\theta )-W_{\psi ,N_{1},N_{2}}(\tau
)\right) ^{2}}{\varepsilon ^{2}}= \\
&=&\frac{\mathsf{E}\,\int_{\tau }^{\tau +\theta }V_{\psi ,N_{1},N_{2}}(s)\,ds%
}{\varepsilon ^{2}}\,\leq \\
&\leq &\frac{\theta \cdot \left( C_{12}(\alpha _{12},\psi
_{1})/N_{1}+C_{21}(\alpha _{21},\psi _{2})/N_{2}\right) }{\varepsilon ^{2}}\,
\end{eqnarray*}%
Using this estimate one can check the sufficient condition of Aldous. Then
Proposition~\ref{e-plot-Fpsi} is proved.

This concludes also the proof of Proposition~\ref{p:family-dense}.

\subsubsection{Weak solutions}

\begin{definition}
We say that the pair of functions $M(t)=(m_{1}(t,x),m_{2}(t,x))$ is a weak
solution of the system~(\ref{cont_1})-(\ref{cont_2}), if for any pair $\phi
_{1}(x),\phi _{2}(x)\in S(\R )$ the following identities hold%
\begin{eqnarray*}
\left\langle M(t),\phi \right\rangle & = & \left\langle M(0),\phi
\right\rangle + \\
& & +\int _{0}^{t}\left\langle M(s),\left(v_{1}\phi _{1}^{\prime }-\alpha
_{12}\phi _{1}+\alpha _{21}\phi _{2},v_{2}\phi _{2}^{\prime }-\alpha
_{12}\phi _{1}+\alpha _{21}\phi _{2}\right)\right\rangle ds,
\end{eqnarray*}
where $\phi (x)=\left(\phi _{1}(x),\phi _{2}(x)\right)$, and the action of $%
G(x)=(g_{1}(x),g_{2}(x))$ on the test function $\phi (x)$ can be written as%
\[
\left\langle G,\phi \right\rangle =\int g_{1}(x)\phi _{1}(x)\, dx+\int
g_{2}(x)\phi _{2}(x)\, dx
\]
\end{definition}

Note that from the representation~(\ref{eq:W-repr}) and the identity~(\ref%
{eq:LF-action}) it follows that%
\begin{eqnarray*}
\left\langle M_{N_{1},N_{2}}(t),\phi \right\rangle & = & W_{\phi
,N_{1},N_{2}}(t)+\left\langle M_{N_{1},N_{2}}(0),\phi \right\rangle + \\
& & +\int _{0}^{t}\left\langle M_{N_{1},N_{2}}(s),\left(v_{1}\phi
_{1}^{\prime }-\alpha _{12}\phi _{1}+\alpha _{21}\phi _{2},v_{2}\phi
_{2}^{\prime }-\alpha _{12}\phi _{1}+\alpha _{21}\phi
_{2}\right)\right\rangle ds,
\end{eqnarray*}
Let $h=h(t)\in \Pi ^{T}=D([0,T],S^{\prime }(\R )\times S^{\prime }(\R ))$. For
fixed $\phi $ define the functional%
\[
J_{\phi ,T}(h)=\sup _{t\leq T}\left|\left\langle h(t),\phi \right\rangle
-\left\langle h(0),\phi \right\rangle -\int _{0}^{t}\left\langle
h(s),\left(v_{1}\phi _{1}^{\prime }-\alpha _{12}\phi _{1}+\alpha _{21}\phi
_{2},v_{2}\phi _{2}^{\prime }-\alpha _{12}\phi _{1}+\alpha _{21}\phi
_{2}\right)\right\rangle ds\right|.
\]
In particular,
\[
\sup _{t\leq T}\left|W_{\phi ,N_{1},N_{2}}(t)\right|=J_{\phi
,T}(M_{N_{1},N_{2}}).
\]
The rest of the proof is standard (see~\cite{KipnLand}) and consists of
three steps.

\smallskip {} \noindent \emph{Step 1.} ~From the definition of topology on $%
\Pi ^{T}$ it follows that $J_{\phi ,T}(\cdot ):\,\Pi ^{T}\rightarrow \R _{+}$
is a continuous functional.

\smallskip {} \noindent \emph{Step 2.~} Note that
\[
\forall \varepsilon >0\quad P\left\{ J_{\phi
,T}(M_{N_{1},N_{2}})>\varepsilon \right\} \equiv P_{N_{1},N_{2}}^{T}\left\{
h:\,J_{\phi ,T}(h)>\varepsilon \right\} \rightarrow 0\qquad
(N_{1},N_{2}\rightarrow \infty )
\]%
by Kolmogorov inequality and Corollary~\ref{c-w2}.

\smallskip{} \noindent \emph{Step 3.} As $J_{\phi ,T}(\cdot )$ is
continuous, then the set $\left\{ h:\, J_{\phi ,T}(h)>0\right\} $ is open in
$\Pi ^{T}$. It follows now that for any limiting point $P_{\infty }^{T}$ of
the family $\left\{ P_{N_{1},N_{2}}^{T}\right\} _{N_{1},N_{2}}$ we have
\[
P_{\infty }^{T}\left\{ h:\, J_{\phi ,T}(h)>\varepsilon \right\} \leq \limsup
_{N_{1},N_{2}}P_{N_{1},N_{2}}^{T}\left\{ h:\, J_{\phi ,T}(h)>\varepsilon
\right\} .
\]
That is, for any $\varepsilon >0$ we have $P_{\infty }^{T}\left\{ h:\,
J_{\phi ,T}(h)>\varepsilon \right\} =0$. In other words, all limiting points
$P_{\infty }^{T}$ of the family $\left\{ P_{N_{1},N_{2}}^{T}\right\}
_{N_{1},N_{2}}$ have support on the set $\left\{ h:\, J_{\phi
,T}(h)=0\right\} $, which consists of weak solutions of~(\ref{cont_1})-(\ref%
{cont_2}).

This completes proof of Proposition~\ref{p-Limit-points}.

\medskip{} The problem of uniqueness of the weak solution of~(\ref{cont_1})-(%
\ref{cont_2}) is quite simple because the system~(\ref{cont_1})-(\ref{cont_2}%
) is \emph{linear}. In the Subsection~\ref{sub:Continuous-model} we shall
see that this system of first order differential equations has a unique
classical solution which can be obtained in explicit way.

\subsection{Time asymptotics for the continuous model}

We prove here Theorem~\ref{t-a-d-cont}.\label{sub:Continuous-model}

Define the means (mass centrum) $a_{i}(t)=\displaystyle \int xm_{i}(t,x)\,
dx $ and variance (momentum of inertia) $d_{i}(t)=\displaystyle \int
(x-a_{i}(t))^{2}m_{i}(t,x)\, dx\, $.

From (\ref{cont_1})--(\ref{cont_2}) we get the following equations for the
means
\begin{eqnarray*}
\dot{a_{1}} &=&v_{1}+\alpha _{12}\left( a_{2}-a_{1}\right) \\
\dot{a_{2}} &=&v_{2}+\alpha _{21}\left( a_{1}-a_{2}\right)
\end{eqnarray*}%
It follows that equation for $a_{2}(t)-a_{1}(t)$ is closed and has the
following solution%
\[
a_{2}(t)-a_{1}(t)=\frac{v_{2}-v_{1}}{\alpha _{12}+\alpha _{21}}\left(
1-e^{-(\alpha _{12}+\alpha _{21})t}\right) +\left( a_{2}(0)-a_{1}(0)\right)
e^{-(\alpha _{12}+\alpha _{21})t}.
\]%
Thus%
\[
{a_{2}(t)-a_{1}(t)\rightarrow \frac{v_{2}-v_{1}}{\alpha _{12}+\alpha _{21}}}%
\quad \quad (t\rightarrow +\infty )
\]%
and\noindent ~similarly%
\[
{\frac{d}{dt}a_{i}(t)\rightarrow \frac{\alpha _{21}v_{1}+\alpha _{12}v_{2}}{%
\alpha _{12}+\alpha _{21}}}\quad \quad (t\rightarrow +\infty )
\]

~

The equations for variances are%
\begin{eqnarray*}
\dot{d_{1}} &=&\alpha _{12}\left( d_{2}-d_{1}\right) +\alpha _{12}\left(
a_{2}(t)-a_{1}(t)\right) ^{2} \\
\dot{d_{2}} &=&\alpha _{21}\left( d_{1}-d_{2}\right) +\alpha _{21}\left(
a_{1}(t)-a_{2}(t)\right) ^{2}
\end{eqnarray*}%
Or, equivalently%
\begin{eqnarray*}
\frac{d}{dt}\left( \alpha _{21}d_{1}+\alpha _{12}d_{2}\right) &=&2\alpha
_{12}\alpha _{21}\left( a_{2}(t)-a_{1}(t)\right) ^{2} \\
\frac{d}{dt}\left( d_{2}-d_{1}\right) &=&-\left( \alpha _{12}+\alpha
_{21}\right) \left( d_{2}-d_{1}\right) +\left( \alpha _{21}-\alpha
_{12}\right) \left( a_{2}(t)-a_{1}(t)\right) ^{2}
\end{eqnarray*}%
From this we get%
\[
d_{2}(t)-d_{1}(t)\rightarrow {const}=\left( \frac{v_{2}-v_{1}}{\alpha
_{12}+\alpha _{21}}\right) ^{2}\cdot \frac{\alpha _{21}-\alpha _{12}}{\alpha
_{12}+\alpha _{21}}\newline
\]%
and%
\[
\noindent \frac{d}{dt}\left( \alpha _{21}d_{1}+\alpha _{12}d_{2}\right)
\rightarrow 2\alpha _{12}\alpha _{21}\left( \frac{v_{2}-v_{1}}{\alpha
_{12}+\alpha _{21}}\right) ^{2}\newline
\]%
~

Thus the growth of variances is asymptotically linear. Moreover, both are
asymptotically equal.

Now we come to the solution of the equations. Define the Fourier transforms%
\[
m_{i}(x,t)=\int \exp (ixp)g_{i}(p,t)dp
\]%
We get%
\[
\frac{\partial g_{1}}{\partial t}+v_{1}ipg_{1}=\alpha _{12}(g_{2}-g_{1})
\]%
\[
\frac{\partial g_{2}}{\partial t}+v_{2}ipg_{2}=\alpha _{21}(g_{1}-g_{2})
\]%
with initial conditions $m_{i}(0,x)=m_{i}(x)$, $i=1,2$. We write this system
in the vector form%
\[
\frac{dg}{dt}=Ag
\]%
where%
\[
A=\left(
\begin{array}{cc}
-iv_{1}p-\alpha _{12} & \alpha _{12} \\
\alpha _{21} & -iv_{2}p-\alpha _{21}%
\end{array}%
\right)
\]%
For eigenvalues we have%
\[
\lambda _{\pm }=-\frac{a}{2}\pm \sqrt{\frac{a^{2}}{4}-b}
\]%
where%
\[
a=i(v_{1}+v_{2})p+\alpha _{12}+\alpha _{21},\qquad
b=-v_{1}v_{2}p^{2}+ip(v_{1}\alpha _{21}+v_{2}\alpha _{12})
\]%
One can write the solution as%
\[
g=C_{+}\phi _{+}\exp (t\lambda _{+})+C_{-}\phi _{-}\exp (t\lambda _{-})
\]%
where $\phi _{\pm }$ are eigenfunctions. Note that for small $p$ there are
two roots. One has $\mathrm{Re}\,\lambda _{-}<0$, thus strongly decreasing term.
Another is%
\begin{equation}
\lambda _{+}=c_{1}p+c_{2}p^{2}+O(p^{3}),c_{2}\neq 0  \label{eigenv}
\end{equation}%
for small $p$.

Let $\xi _{t}$ be a random variable with density $m(x,t)$, $g(k)$ - its
characteristic function. We are interested in $\frac{1}{\sqrt{t}}(\xi
_{t}-a),a=\sE\xi _{t}$, its characteristic function is%
\[
\exp (-ia\frac{k}{\sqrt{t}})g(\frac{k}{\sqrt{t}})
\]%
Using (\ref{eigenv}) we get the result.

\begin{remark}
One can see that there is no solution of the type%
\[
m_{i}(t,x)=f_{i}(x-vt)
\]%
as then $f_{i}$ would be exponents.
\end{remark}

\begin{remark}
For the singular initial conditions, that is when $x_{k}^{(i)}(0)=0$ for $%
k=1,\ldots,N_{i}$; $i=1,2$, one can get the same asymptotic results.
\end{remark}

\section{Appendix}

\subsection{Probability measures on the Skorohod space: tightness}

Let $\left\{ \left(\xi _{t}^{n},t\in [0,T]\right)\right\} _{n\in \mathbf{N}}$
\ be a sequence of real random processes which trajectories are
right-continuous and admit left-hand limits for every $0<t\leq T$ . We will
consider $\xi ^{n}$ as random elements with values in the Skorohod space $%
D_{T}(\R ):=D\left([0,T],\R ^{1}\right)$ with the standard topology. Denote $%
P_{T}^{n}$ the distribution of $\xi ^{n}$, defined on the measurable space $%
\left(D_{T}(\R ),\mathcal{B}\left(D_{T}(\R )\right)\right)$. The following
result can be found in~\cite{billingsley}.

\begin{theorem}
~\ The sequence of probability measures $\left\{ P_{T}^{n}\right\} _{n\in
\mathbf{N}}$ is tight iff the following two conditions hold:

1) for any $\varepsilon >0$ there is $\, C(\varepsilon )>0$ such that%
\[
\sup _{n}P_{T}^{n}\left(\sup _{0\leq t\leq T}\left|\xi
_{t}^{n}\right|>C(\varepsilon )\right)\leq \varepsilon \, ;
\]

2) for any $\varepsilon >0$%
\[
\lim _{\gamma \rightarrow 0}\limsup _{n}P_{T}^{n}\left(\xi _{\cdot }:\,
w^{\prime }(\xi ;\gamma )>\varepsilon \right)=0\, ,
\]
where for any function $f:\, [0,T]\rightarrow \R $ and any $\gamma >0$ we
define%
\[
w^{\prime }(f;\gamma )=\inf _{\left\{ t_{i}\right\} _{i=1}^{r}}\, \max
_{i<r}\, \sup _{t_{i}\leq s<t<t_{i+1}}\left|f(t)-f(s)\right|\, ,
\]
moreover the $\inf $ is over all partitions of the interval $[0,T]$ such
that
\[
0=t_{0}<t_{1}<\cdots <t_{r}=T,\qquad t_{i}-t_{i-1}>\gamma ,\quad i=1,\ldots
,r.
\]
\end{theorem}

The following theorem is known as the sufficient condition of Aldous~\cite%
{KipnLand}.

\begin{theorem}
Condition 2) of the previous theorem follows from the following condition%
\[
\forall \varepsilon >0\qquad \lim _{\gamma \rightarrow 0}\, \limsup _{n}\,
\sup _{\tau \in \mathcal{R}_{T},\, \theta \leq \gamma }P_{T}^{n}\left(\left|\xi _{\tau
+\theta }-\xi _{\tau }\right|>\varepsilon \right)=0\, ,
\]
where $\mathcal{R}_{T}$ \ is the set of Markov moments (stopping times) not
exceeding $T$ .
\end{theorem}

\subsection{Strong topology on the Skorohod space. Mitoma theorem}

\label{sub:Strong-topology-on}Remind that Schwartz space $S(\R )$ is a Frechet
space (complete locally convex space, the topology of which is generated by
countable family of seminorms, that implies metrizability, see~\cite%
{ReedSimon}). In the dual space $S^{\prime }(\R )$ of tempered distributions
there are at least two ways to define topology (both not metrizable):

1) \emph{weak topology} on $S^{\prime }(\R )$, where all functionals
\[
\left\langle \, \cdot \, ,\phi \right\rangle ,\quad \phi \in S(\R )
\]
are continuous.

2) \emph{strong topology} on $S^{\prime }(\R )$, which is generated by the set
of seminorms%
\[
\left\{ \rho _{A}(M)=\sup _{\phi \in A}\left|\left\langle M,\phi
\right\rangle \right|\, :\, \, A\subset S(\R )\, -\, {bounded}\right\} .
\]
We shall consider $S^{\prime }(\R )$ as equipped with the strong topology.
Details can be found in~\cite{ReedSimon}.

The problem of introducing of the Skorohod topology on the space $%
D_{T}(S^{\prime}):=D([0,T],S^{\prime}(\R ))$ was studied in~\cite{mitoma} and~%
\cite{Jakub}. The topology on this space is defined as follows. Let $\left\{
\rho _{A}\right\} $ be a family of seminorms, which generates strong
topology in $S^{\prime }(\R )$. For each seminorm $\rho _{A}$ define a
pseudometrics%
\[
d_{A}(y,z)=\inf _{\lambda \in \Lambda }\left\{ \sup
_{t}\left|y_{t}-z_{\lambda (t)}\right|+\sup _{t\not =s}\left|\log \frac{%
\lambda (t)-\lambda (s)}{t-s}\right|\right\} ,\quad y,z\in
D_{T}(S^{\prime}),
\]
where the inf is over the set $\Lambda =\left\{ \lambda =\lambda (t),\, t\in
\lbrack 0,T]\right\} $ of all strictly increasing maps of the interval $%
[0,T] $ into itself. Equipped with the topology of the projective limit for
the family $\left\{ d_{A}\right\} $ the set $D_{T}(S^{\prime})$ becomes a
completely regular topological space.

Let $\mathcal{B}(D_{T}(S^{\prime}))$ be the corresponding Borel $\sigma $%
-algebra. Let $\left\{ P_{n}\right\} $ be a sequence of probability measures
on $\left(D_{T}(S^{\prime}),\mathcal{B}(D_{T}(S^{\prime}))\right)$. For each
$\phi \in S(\R )$ consider a map $\mathcal{I} _{\phi }:\, y\in
D_{T}(S^{\prime})\rightarrow y_{\cdot }(\phi )\in D_{T}(\R )$. The following
result belongs to I.~Mitoma~\cite{mitoma}.

\begin{theorem}

Suppose that for any $\phi \in S(\R )$ the sequence $\left\{ P_{n}\mathcal{I}
_{\phi }^{-1}\right\} $ is tight in in $D_{T}(\R )$. Then the sequence~$%
\left\{ P_{n}\right\} $ itself is tight in~$D_{T}(S^{\prime})$.
\end{theorem}

\end{document}